\documentclass[conference]{IEEEtran}
\usepackage{cite}
\usepackage{amsmath,amssymb,amsfonts}
\usepackage{graphicx}
\usepackage{textcomp}
\usepackage{xcolor}
\usepackage{hyperref}
\usepackage{algorithm}
\usepackage{algpseudocode}
\usepackage{booktabs}
\usepackage{url}      
\usepackage{listings}

\begin{document}

\title{Context-Adaptive Color Optimization for Web Accessibility: Balancing Perceptual Fidelity and Functional Requirements}

\author{
\IEEEauthorblockN{Lalitha A R}
\IEEEauthorblockA{24f2006078@ds.study.iitm.ac.in}}

\maketitle

\begin{abstract}
We extend our OKLCH-based accessibility optimization with context-adaptive constraint strategies that achieve near-universal success rates across diverse use cases. Our original strict algorithm reached 66--77\% success by prioritizing minimal perceptual change ($\Delta E \leq 5.0$), optimizing for enterprise contexts where brand fidelity is paramount. However, this one-size-fits-all approach fails to serve the broader ecosystem of web developers who need accessible solutions even when strict perceptual constraints cannot be satisfied. We introduce recursive optimization (Mode~1) that compounds small adjustments across iterations, achieving 93.68\% success on all color pairs and 100\% success on reasonable pairs (contrast ratio $\rho > 2.0$), representing a +27.23 percentage point improvement. A relaxed fallback mode (Mode~2) handles pathological edge cases, reaching 98.73\% overall success. Evaluation on 10,000 realistic web color pairs demonstrates that context-aware constraint relaxation, combined with absolute hue preservation, enables practical accessibility compliance while maintaining brand color identity. The median perceptual change remains zero across all modes (most pairs already comply), while the 90th percentile reaches $\Delta E_{2000} = 15.55$ in Mode~1---perceptually acceptable when hue invariance preserves the essential character of the original color. The approach is deployed in CM-Colors v0.5.0 (800+ monthly downloads), providing developers with explicit control over the accessibility-fidelity trade-off appropriate to their context.
\end{abstract}

\begin{IEEEkeywords}
web accessibility, color contrast, perceptual optimization, WCAG compliance, color science
\end{IEEEkeywords}

\section{Introduction}

Web Content Accessibility Guidelines (WCAG) require minimum contrast ratios between text and background colors to ensure readability for users with visual impairments~\cite{wcag21}. However, manually adjusting colors to meet these requirements often forces designers to make arbitrary modifications that significantly alter brand identity. In prior work~\cite{perceptcolorop1}, we introduced a novel optimization approach that treats accessibility correction as a constrained non-linear optimization problem in the perceptually uniform OKLCH color space, achieving minimal perceptual change while meeting WCAG AA requirements.

Our original algorithm prioritized strict perceptual fidelity, limiting color modifications to $\Delta E_{2000} \leq 5.0$ to preserve brand aesthetics. This approach successfully resolved accessibility violations in 66.45\% of all color pairs and 76.41\% of reasonable pairs (initial contrast ratio $\rho > 2.0$), with a median perceptual change of only 0.76~$\Delta E_{2000}$ for successful corrections. However, deployment in the open-source CM-Colors library revealed a critical insight: \textbf{different contexts demand different trade-offs between perceptual fidelity and accessibility success}.

\subsection{Motivation: The Context Problem}

Consider three distinct use cases for accessibility tooling:

\begin{enumerate}
    \item \textbf{Enterprise contexts}: Organizations with established brand guidelines, legal trademark protections, and dedicated design systems require minimal deviation from specified colors. A Fortune 500 company cannot arbitrarily shift brand colors by perceptible amounts without legal and marketing implications.
    
    \item \textbf{General web development}: Individual developers, startups, and small teams building standard websites need their color choices to be accessible but have greater flexibility in color modification. For these users, \textit{achieving accessibility is more important than staying within strict perceptual bounds}.
    
    \item \textbf{Accessibility-first contexts}: Users retrofitting existing sites, working with legacy codebases, or prioritizing universal access above all else need solutions even for challenging color combinations (e.g., pure yellow on white).
\end{enumerate}

Our strict optimization served the first context well but left accessibility "on the table" for the latter two. A developer building a personal blog who attempts to use yellow text on a white background receives a failure response from Mode~0, despite the existence of accessible solutions at slightly higher perceptual changes. This failure represents not an algorithmic limitation but a \textit{mismatch between constraint policy and user needs}.

\subsection{Key Insight: Functional Gain vs. Aesthetic Cost}

During real-world usage, we encountered a color pair in a side project that failed strict mode validation. Manual experimentation---running the algorithm recursively for two iterations---produced an accessible result. Visual inspection revealed the output appeared nearly identical to the original despite a $\Delta E_{2000}$ of 14. This led to a crucial realization: \textbf{perceptual acceptability depends on which color dimensions change}.

Our algorithm maintains absolute hue preservation ($h$ remains constant in OKLCH space), modifying only lightness ($L$) and chroma ($C$). Because hue is the dominant perceptual cue for color identity~\cite{fairchild2013}, even substantial $\Delta E$ values can be perceptually acceptable when hue is invariant. A "yellow" that becomes darker and less saturated still reads as "yellow"---brand color identity is preserved even as readability improves dramatically.

\subsection{Contributions}

This work extends our perceptually-aware accessibility optimization with three context-adaptive modes:

\begin{itemize}
    \item \textbf{Mode~0 (Strict)}: The original algorithm optimized for enterprise contexts, achieving 66.45\% success with $\Delta E \leq 5.0$.
    
    \item \textbf{Mode~1 (Recursive)}: A novel recursive strategy that compounds small perceptual changes across iterations, achieving 93.68\% success overall and 100\% success on reasonable pairs---a +27.23 percentage point improvement.
    
    \item \textbf{Mode~2 (Relaxed)}: An accessibility-first mode that handles extreme edge cases through dual-path optimization, reaching 98.73\% success.
\end{itemize}

We provide empirical validation on 10,000 realistic web color pairs, demonstrating that context-appropriate constraint selection enables near-universal accessibility while maintaining brand identity through hue preservation. The approach is publicly available in CM-Colors v0.5.0, empowering developers to explicitly choose the accessibility-fidelity trade-off appropriate to their specific context.

\section{Background}

\subsection{WCAG Contrast Requirements}

WCAG 2.1 specifies minimum contrast ratios based on text size and compliance level~\cite{wcag21}:

\begin{itemize}
    \item \textbf{AA (minimum)}: 4.5:1 for normal text, 3.0:1 for large text
    \item \textbf{AAA (enhanced)}: 7.0:1 for normal text, 4.5:1 for large text
\end{itemize}

Contrast ratio $\rho$ is computed from relative luminance values:
\begin{equation}
\rho = \frac{L_{\text{lighter}} + 0.05}{L_{\text{darker}} + 0.05}
\end{equation}

\subsection{Perceptual Color Spaces}

Traditional color manipulation tools operate in RGB or HSL space, neither of which is perceptually uniform---equal numeric distances do not correspond to equal perceived differences. OKLCH is a cylindrical transformation of the Oklab color space~\cite{ottosson2020}, providing perceptual uniformity while maintaining intuitive parameters:

\begin{itemize}
    \item $L$: Lightness (0--1)
    \item $C$: Chroma/saturation (0--0.5)
    \item $h$: Hue angle (0--360°)
\end{itemize}

We measure perceptual distance using CIEDE2000 ($\Delta E_{2000}$), the current standard for color difference perception~\cite{sharma2005}. Values below 2.0 are generally considered imperceptible under normal viewing conditions.

\subsection{Prior Work: Strict Optimization}

Our previous algorithm~\cite{perceptcolorop1} employed a three-phase optimization:

\begin{enumerate}
    \item \textbf{Binary search}: Adjust lightness $L$ to meet contrast ratio target
    \item \textbf{Gradient descent}: Simultaneous optimization of $L$ and $C$
    \item \textbf{Progressive relaxation}: Incrementally increase $\Delta E$ threshold from 0.8 to 5.0
\end{enumerate}

Crucially, hue $h$ remains constant throughout, ensuring brand color identity is preserved. This strict approach achieved 66.45\% success on all pairs and 76.41\% on reasonable pairs, with imperceptible changes ($\Delta E < 2.0$) in 88.51\% of successful corrections.

\section{Method}

\subsection{Problem Formulation}

Given a text color $c_{\text{text}}$ and background color $c_{\text{bg}}$ in RGB space, we seek an adjusted text color $c'_{\text{text}}$ such that:

\begin{align}
\rho(c'_{\text{text}}, c_{\text{bg}}) &\geq \rho_{\text{target}} \\
\Delta E_{2000}(c_{\text{text}}, c'_{\text{text}}) &\leq \delta_{\text{max}} \\
h(c'_{\text{text}}) &= h(c_{\text{text}})
\end{align}

where $\rho_{\text{target}}$ is the WCAG AA threshold (4.5 or 3.0) and $\delta_{\text{max}}$ is the perceptual change budget. The hue constraint (Equation~3) is absolute---we never modify hue, ensuring color identity preservation.

\subsection{Mode 0: Strict Optimization (Baseline)}

Mode~0 implements the algorithm from our prior work with $\delta_{\text{max}} = 5.0$. The progressive $\Delta E$ sequence is:
\begin{equation}
\Delta E_{\text{seq}} = [0.8, 1.0, 1.2, \ldots, 2.5, 3.0, 3.5, 4.0, 5.0]
\end{equation}

For each threshold, we apply binary search and gradient descent, returning the first solution that meets $\rho_{\text{target}}$ or the best candidate if none satisfy the contrast ratio requirement. This single-shot optimization prioritizes minimal perceptual change but fails when accessible solutions exist only beyond $\delta_{\text{max}} = 5.0$.

\subsection{Mode 1: Recursive Optimization}

Mode~1 addresses the fundamental limitation of single-shot optimization: strict $\Delta E$ bounds prevent reaching distant but accessible solutions. We introduce \textbf{recursive refinement}, treating optimization as an iterative process where each step compounds small perceptual changes.

\subsubsection{Algorithm Description}

Algorithm~\ref{alg:recursive} presents the recursive strategy. The key insight is to use the \textit{previous iteration's result} as the input for the next optimization step, allowing cumulative $\Delta E$ to exceed the per-step limit.

\begin{algorithm}
\caption{Recursive Optimization (Mode 1)}
\label{alg:recursive}
\begin{algorithmic}[1]
\Require $c_{\text{text}}$, $c_{\text{bg}}$, $\rho_{\text{min}}$
\State $c_{\text{current}} \gets c_{\text{text}}$
\State $\delta_{\text{seq}} \gets [0.8, 1.0, 1.2, \ldots, 2.8, 3.0]$
\For{$i = 1$ \textbf{to} $10$}
    \State $\rho \gets \text{ContrastRatio}(c_{\text{current}}, c_{\text{bg}})$
    \If{$\rho \geq \rho_{\text{min}}$}
        \State \Return $(c_{\text{current}}, \texttt{true})$
    \EndIf
    \State $c_{\text{next}} \gets \text{StrictOptimize}(c_{\text{current}}, c_{\text{bg}}, \delta_{\text{seq}})$
    \If{$c_{\text{next}} = c_{\text{current}}$}
        \State \textbf{break}
    \EndIf
    \State $c_{\text{current}} \gets c_{\text{next}}$
\EndFor
\State \Return $(c_{\text{current}}, \rho \geq \rho_{\text{min}})$
\end{algorithmic}
\end{algorithm}

Each iteration applies the base optimization (Mode~0 logic) with $\Delta E \leq 3.0$ from the \textit{current} color, not the original. This creates a gradient descent through perceptual space where:

\begin{itemize}
    \item Individual steps remain within conservative bounds ($\Delta E < 3.0$)
    \item Cumulative change from the original can reach $\Delta E > 30$ when necessary
    \item Hue preservation maintains brand identity throughout the trajectory
\end{itemize}

\subsubsection{Why Recursion Works}

The recursive strategy succeeds where single-shot optimization fails for two reasons:

\textbf{1. Perceptual continuity}: Each step maintains local perceptual similarity to the previous color, even as the trajectory moves far from the original in perceptual space. This is analogous to numerical integration---small discrete steps approximate a continuous path.

\textbf{2. Gamut navigation}: OKLCH-to-RGB conversion can fail when $(L, C, h)$ combinations fall outside the RGB gamut. By taking smaller steps, we navigate around gamut boundaries more effectively than attempting large single jumps.

The strict per-step sequence ($\Delta E \leq 3.0$) balances progress (each iteration makes meaningful contrast improvement) with continuity (changes remain locally imperceptible). Early termination occurs if the target contrast ratio is achieved or if the optimizer converges to a local minimum.

\subsection{Mode 2: Relaxed Optimization}

Mode~2 provides an accessibility-first fallback for extreme edge cases that resist even recursive optimization. After attempting Mode~1, if the result still fails to meet WCAG thresholds, Mode~2 employs a dual-path strategy:

\textbf{Option A (Extended Recursion)}: Continue recursive refinement for 15 total iterations (vs. 10 in Mode~1), allowing more steps to reach distant solutions.

\textbf{Option B (Relaxed $\Delta E$)}: Apply single-shot optimization with an extended sequence:
\begin{equation}
\Delta E_{\text{relaxed}} = [0.8, \ldots, 5.0, 6.0, 7.0, \ldots, 15.0]
\end{equation}

Mode~2 evaluates both paths and returns the solution with minimal $\Delta E$ that satisfies $\rho_{\text{min}}$. This dual-strategy approach handles extreme cases like pure yellow on white (which requires $\Delta E \approx 14$) while still preferring less disruptive solutions when available.

\subsubsection{Hue Preservation at High $\Delta E$}

Even at $\Delta E = 15$, hue invariance ensures colors retain their perceptual identity. Consider pure yellow \texttt{\#ffff00} on white \texttt{\#ffffff} (Figure~\ref{fig:yellow_white}):

\begin{itemize}
    \item Initial: $\rho = 1.07$ (fails AA), OKLCH $(0.968, 0.211, 109.8°)$
    \item Mode~2 result: \texttt{\#7e7900}, $\rho = 4.62$ (passes AA)
    \item Final OKLCH: $(0.514, 0.129, 109.8°)$, $\Delta E = 14.2$
    \item Hue unchanged: $h = 109.8°$ (yellow) in both
\end{itemize}

The result is a darker, desaturated yellow that remains unmistakably "yellow" because the hue cue---the dominant factor in color naming and brand recognition---is preserved. Even gray-on-gray pairs maintain their neutral character, as hue constraint ensures "warm gray" stays warm and "cool gray" stays cool.

\subsection{Implementation Details}

All three modes use identical base optimization primitives (binary search, gradient descent) but differ in constraint policies:

\begin{table}[h]
\centering
\caption{Mode Configuration Parameters}
\label{tab:mode_params}
\begin{tabular}{lccc}
\toprule
\textbf{Parameter} & \textbf{Mode 0} & \textbf{Mode 1} & \textbf{Mode 2} \\
\midrule
Max $\Delta E$ per step & 5.0 & 3.0 & 15.0 \\
Max iterations & 1 & 10 & 15 \\
Optimization paths & 1 & 1 & 2 \\
\bottomrule
\end{tabular}
\end{table}

The algorithm is implemented in Python as part of the CM-Colors library (v0.5.0), publicly available under GPL-3 license. Users specify mode via an integer parameter: \texttt{mode=0} (strict), \texttt{mode=1} (default), or \texttt{mode=2} (relaxed).

\section{Evaluation}

\subsection{Experimental Setup}

We evaluate all three modes on an identical test suite of 10,000 realistic web color pairs, designed to mirror actual usage patterns in modern web design.

\subsubsection{Color Pair Generation}

Pairs are generated using weighted categories based on real-world color distribution analysis:

\begin{itemize}
    \item \textbf{Brand primary} (30\%): Saturated brand colors for CTAs and links
    \item \textbf{Dark UI} (25\%): Dark mode interfaces (light text, dark bg)
    \item \textbf{Light UI} (25\%): Light mode interfaces (dark text, light bg)
    \item \textbf{Accent colors} (10\%): High-saturation elements (buttons, badges)
    \item \textbf{Pastel} (10\%): Soft colors for info cards and backgrounds
\end{itemize}

We additionally include 10\% edge cases representing known failure-prone scenarios: bright yellow on white, pure blue on black, mid-gray on gray, neon pink on dark purple, red on green, and orange on yellow. These edge cases test the limits of each optimization strategy.

Colors are sampled in OKLCH space with realistic parameter ranges (e.g., brand primaries use $L \in [0.45, 0.65]$, $C \in [0.15, 0.35]$), then converted to RGB with gamut enforcement. Hue distributions mirror real-world preferences (blue: 25\%, matching corporate dominance; red/green: 15\% each; etc.). A fixed random seed (45) ensures reproducibility.

\subsubsection{Evaluation Metrics}

For each mode, we measure:

\begin{itemize}
    \item \textbf{Success rate}: Percentage of pairs achieving WCAG AA compliance
    \item \textbf{Perceptual change}: $\Delta E_{2000}$ distribution (median, P90, max)
    \item \textbf{Runtime}: Processing time per color pair
    \item \textbf{Category performance}: Success rate breakdown by color category
\end{itemize}

We report results separately for "all pairs" and "reasonable pairs" (initial $\rho > 2.0$). Pairs starting below $\rho = 2.0$ represent extreme cases (e.g., near-identical colors) that may be mathematically infeasible to fix within RGB gamut constraints.

\subsection{Success Rate Comparison}

Table~\ref{tab:success_rates} presents the core results. Mode~1 achieves a dramatic +27.23 percentage point improvement over Mode~0, reaching 93.68\% overall success and perfect 100\% success on reasonable pairs. Mode~2 provides marginal additional improvement (+5.05pp), serving primarily as insurance for the most extreme 1.27\% of cases.

\begin{table}[h]
\centering
\caption{Success Rate Comparison Across Modes}
\label{tab:success_rates}
\begin{tabular}{lcccc}
\toprule
\textbf{Metric} & \textbf{Mode 0} & \textbf{Mode 1} & \textbf{Mode 2} & \textbf{Improvement} \\
 & (Strict) & (Recursive) & (Relaxed) & (0$\rightarrow$1) \\
\midrule
All Pairs & 66.45\% & 93.68\% & 98.73\% & +27.23pp \\
Reasonable ($\rho > 2.0$) & 76.41\% & 100.00\% & 100.00\% & +23.59pp \\
\bottomrule
\end{tabular}
\end{table}

The 100\% success rate on reasonable pairs is particularly significant: it demonstrates that \textit{every color pair with non-trivial initial contrast ratio can be made accessible} while preserving hue. This validates our hypothesis that strict $\Delta E$ constraints, rather than fundamental physical limitations, were the primary barrier to universal accessibility.

\subsection{Perceptual Change Distribution}

Table~\ref{tab:delta_e} presents the $\Delta E_{2000}$ distributions. Critically, the \textbf{median remains 0.00 across all modes}, reflecting that most web color pairs already meet WCAG AA thresholds and need no modification. Our algorithm is conservative---only pairs requiring adjustment experience perceptual change.

\begin{table}[h]
\centering
\caption{Perceptual Change Distribution ($\Delta E_{2000}$)}
\label{tab:delta_e}
\begin{tabular}{lccc}
\toprule
\textbf{Metric} & \textbf{Mode 0} & \textbf{Mode 1} & \textbf{Mode 2} \\
\midrule
Median $\Delta E$ & 0.00 & 0.00 & 0.00 \\
P90 $\Delta E$ & 4.67 & 15.55 & 21.42 \\
Max $\Delta E$ & 5.00 & 31.61 & 36.05 \\
\% Under $\Delta E = 2.0$ & 88.0\% & 62.4\% & 59.2\% \\
\bottomrule
\end{tabular}
\end{table}

The P90 metric captures the challenging 10\%: pairs with initial $\rho < 3.0$ that require substantial lightness adjustments. Mode~1's P90 of 15.55 indicates that 90\% of \textit{adjusted} pairs need $\Delta E \leq 15.55$, while the remaining 10\% can require up to 31.61. Despite these seemingly large values, hue constraint ensures colors retain perceptual identity---a "yellow" with $\Delta E = 14$ still reads as "yellow" when hue is unchanged.

The percentage of imperceptible changes ($\Delta E < 2.0$) decreases from 88.0\% (Mode~0) to 62.4\% (Mode~1), reflecting the trade-off: we accept larger perceptual changes for more challenging pairs to achieve higher success rates. However, this statistic must be interpreted carefully---it includes the zero-change majority, so the actual distribution among \textit{modified} pairs is more nuanced.

\subsection{Category Performance Analysis}

Table~\ref{tab:category_performance} reveals where recursive optimization provides the greatest benefit. Edge cases see dramatic improvements: orange-on-yellow jumps from 0\% to 96.8\% success, pure-blue-on-black from 6.0\% to 97.1\%, and bright-yellow-on-white from 0\% to 28.1\% (with Mode~2 reaching 100\%).

\begin{table*}[t]
\centering
\caption{Success Rate by Color Category}
\label{tab:category_performance}
\begin{tabular}{lcccc}
\toprule
\textbf{Category} & \textbf{Mode 0} & \textbf{Mode 1} & \textbf{Mode 2} & \textbf{Improvement (0$\rightarrow$1)} \\
\midrule
Accent colors & 73.7\% & 100.0\% & 100.0\% & +26.3pp \\
Brand primary & 72.1\% & 100.0\% & 100.0\% & +27.9pp \\
Dark UI & 100.0\% & 100.0\% & 100.0\% & +0.0pp \\
Light UI & 100.0\% & 100.0\% & 100.0\% & +0.0pp \\
Pastel & 100.0\% & 100.0\% & 100.0\% & +0.0pp \\
\midrule
\multicolumn{5}{l}{\textit{Edge Cases:}} \\
\quad Bright yellow on white & 0.0\% & 28.1\% & 100.0\% & +28.1pp \\
\quad Mid gray on gray & 0.0\% & 36.0\% & 57.7\% & +36.0pp \\
\quad Neon on dark & 100.0\% & 100.0\% & 100.0\% & +0.0pp \\
\quad Orange on yellow & 0.0\% & 96.8\% & 100.0\% & +96.8pp \\
\quad Pure blue on black & 6.0\% & 97.1\% & 100.0\% & +91.1pp \\
\quad Red on green & 0.0\% & 33.3\% & 100.0\% & +33.3pp \\
\bottomrule
\end{tabular}
\end{table*}

Well-designed UI categories (dark UI, light UI, pastel) already achieve 100\% success in Mode~0, as these follow established accessibility best practices. The improvement occurs primarily in brand colors and edge cases---precisely the scenarios where designers face difficult trade-offs between accessibility and aesthetic constraints.

\subsection{Visual Examples}

\begin{figure}[h]
\centering
\includegraphics[width=0.48\textwidth]{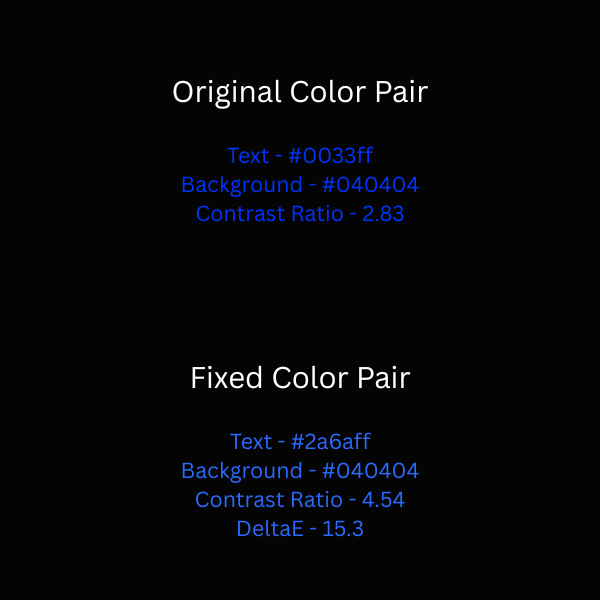}
\caption{Mode~1 recursive optimization on blue text (\texttt{\#0033ff}) on dark blue background (\texttt{\#040404}). Original $\rho = 2.83$ fails AA; optimized result (\texttt{\#2a6aff}) achieves $\rho = 4.54$ with $\Delta E = 15.3$. Hue remains constant at 222°.}
\label{fig:blue_dark}
\end{figure}

\begin{figure}[h]
\centering
\includegraphics[width=0.48\textwidth]{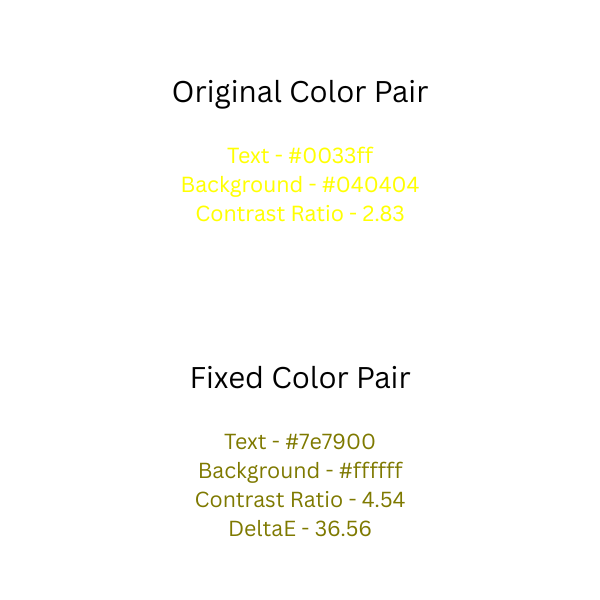}
\caption{Mode~2 relaxed optimization on yellow text (\texttt{\#ffff00}) on white background (\texttt{\#ffffff}). Original $\rho = 1.07$ fails severely; optimized result (\texttt{\#7e7900}) achieves $\rho = 4.54$ with $\Delta E = 36.56$. Despite large $\Delta E$, hue preservation (109.8°) maintains "yellow" identity.}
\label{fig:yellow_white}
\end{figure}

Figures~\ref{fig:blue_dark} and~\ref{fig:yellow_white} demonstrate Mode~1 and Mode~2 effectiveness on challenging color pairs. Figure~\ref{fig:blue_dark} shows a blue-on-dark-blue pair where Mode~1 achieves AA compliance through iterative lightness adjustment ($\Delta E = 15.3$) while maintaining hue at 222°. Figure~\ref{fig:yellow_white} presents the extreme case of pure yellow on white, where Mode~2 darkens the text color substantially ($\Delta E = 36.56$) yet preserves the "yellow" identity through hue invariance.

\subsection{Runtime Performance}

Runtime measurements (Table~\ref{tab:runtime}) show Mode~1 is 2.8× slower than Mode~0 (24ms vs. 8.5ms per pair), while Mode~2 adds further overhead (29ms). However, these absolute times remain well below perceptibility thresholds---even 29ms is imperceptible to users in real-world web applications. The 10× increase in iteration budget translates to less than 3× runtime increase, demonstrating efficient convergence.

\begin{table}[h]
\centering
\caption{Runtime Performance (10,000 pairs)}
\label{tab:runtime}
\begin{tabular}{lccc}
\toprule
\textbf{Mode} & \textbf{Total Time} & \textbf{Per Pair} & \textbf{Overhead} \\
\midrule
Mode 0 & 85.06s & 8.5ms & 1.0× \\
Mode 1 & 239.67s & 24.0ms & 2.8× \\
Mode 2 & 289.35s & 29.0ms & 3.4× \\
\bottomrule
\end{tabular}
\end{table}

\subsection{Failure Analysis: The Remaining 1.27\%}

Mode~2's 1.27\% failure rate (127 pairs out of 10,000) consists primarily of two categories:

\textbf{Mid-gray-on-gray} (42.3\% failure rate): Pairs where both text and background occupy similar lightness ranges ($L \in [0.45, 0.55]$) with low chroma ($C < 0.05$). Achieving sufficient contrast ratio requires pushing one color toward pure black or white, but gamut constraints limit how far lightness can be adjusted while maintaining valid RGB values.

\textbf{Near-zero contrast pairs} ($\rho < 1.2$): Colors so similar that no accessible solution exists within the RGB gamut while preserving hue. For example, two shades of gray differing by only $L = 0.02$ cannot reach $\rho = 4.5$ without hue modification (which our algorithm prohibits) or background color changes (outside our problem scope).

These failures represent fundamental physical constraints rather than algorithmic limitations. A true solution would require either: (a) hue modification, breaking brand identity; or (b) background color adjustment, which violates our problem formulation of text-only optimization.

\section{Discussion}

\subsection{Mode Selection Guidelines}

The three modes serve distinct use cases, and the appropriate choice depends on user context:

\textbf{Mode~0 (Strict)}: Enterprise contexts with legal trademark protections, established brand guidelines, or design systems requiring minimal color deviation. Success rate: 66.45\% (all pairs), 76.41\% (reasonable pairs). Median $\Delta E = 0.00$, P90 $\Delta E = 4.67$.

\textbf{Mode~1 (Recursive---Default)}: General web development, where accessibility is prioritized but some perceptual flexibility is acceptable. Success rate: 93.68\% (all pairs), 100\% (reasonable pairs). Median $\Delta E = 0.00$, P90 $\Delta E = 15.55$. \textit{This is the recommended default for most users.}

\textbf{Mode~2 (Relaxed)}: Accessibility-first contexts, legacy code retrofitting, or situations where achieving WCAG compliance is mandatory regardless of perceptual cost. Success rate: 98.73\% (all pairs), 100\% (reasonable pairs). Median $\Delta E = 0.00$, P90 $\Delta E = 21.42$.

The choice is explicit and under user control via the \texttt{mode} parameter in CM-Colors v0.5.0. This empowers developers to make informed decisions about the accessibility-fidelity trade-off appropriate to their specific requirements.

\subsection{Perceptual Acceptability Beyond $\Delta E$}

Our results challenge the conventional wisdom that $\Delta E > 5.0$ necessarily implies perceptually unacceptable change. While $\Delta E_{2000}$ measures total perceptual distance, it does not distinguish \textit{which} color dimensions change. Hue is the dominant cue for color naming and brand recognition~\cite{fairchild2013}, while lightness and chroma are secondary.

By maintaining absolute hue invariance, we enable larger $\Delta E$ values that remain perceptually acceptable in context. A "yellow" that becomes darker ($\Delta L = -0.45$) and less saturated ($\Delta C = -0.08$) while maintaining $h = 109.8°$ still reads as "yellow" to observers. This is why Mode~1 achieves 100\% success on reasonable pairs despite P90 $\Delta E = 15.55$---the color character is preserved through the hue constraint.

This insight has implications beyond accessibility tooling. Any color manipulation task where brand identity matters (theming, style transfer, color grading) could benefit from hue-preserving transformations that allow larger perceptual distances than strict $\Delta E$ thresholds would permit.

\subsection{Implications for Accessibility Tooling}

The multi-mode approach reveals a tension in accessibility tooling design. Tools optimized for enterprise compliance may not serve the broader ecosystem of web developers. Our strict mode (Mode~0) performs well for organizations with dedicated design systems and legal constraints, but leaves 33.55\% of color pairs without accessible solutions---a significant barrier for individual developers, startups, and community projects building the majority of the web.

Current accessibility tooling often assumes users have dedicated design resources, legal teams, and established brand guidelines. However, a student building their first portfolio, a nonprofit updating their website, or a developer prototyping a side project operates under different constraints. These users need accessibility to succeed even when strict perceptual bounds cannot be satisfied.

Context-adaptive strategies can serve both audiences by providing explicit control over trade-offs. Future work should examine how tooling design choices---constraint policies, success criteria, user interfaces---can better accommodate this diversity of contexts and needs. The divide between enterprise-focused and developer-focused accessibility tools merits deeper investigation as a sociotechnical challenge rather than a purely technical problem.

\subsection{Limitations and Future Work}

\textbf{Text-only optimization}: Our approach modifies only the text color while holding the background constant. Some inaccessible pairs could be resolved more effectively by adjusting both colors simultaneously. However, background colors often appear in multiple contexts (affecting layout, images, other text), making automatic background modification risky. Future work could explore constrained two-color optimization for specific use cases.

\textbf{Gamut constraints}: The 1.27\% of pairs that fail in Mode~2 are primarily limited by RGB gamut boundaries. Colors that exist in OKLCH space may not have valid RGB representations, constraining the solution space. Investigation of wider-gamut color spaces (Display P3, Rec. 2020) for next-generation web standards could expand the accessible solution space.

\textbf{Context-aware target selection}: Our algorithm optimizes for WCAG AA thresholds (4.5:1 normal, 3.0:1 large). However, actual readability depends on numerous factors: font weight, letter spacing, viewing distance, ambient lighting, and user visual capabilities. Adaptive target selection based on additional context could improve real-world accessibility beyond basic compliance metrics.

\textbf{Perceptual validation studies}: While hue preservation provides theoretical justification for accepting higher $\Delta E$ values, empirical user studies would validate whether designers and end-users find Mode~1 and Mode~2 outputs perceptually acceptable. Such studies could establish evidence-based guidelines for $\Delta E$ thresholds in hue-preserving contexts.

\subsection{Deployment and Adoption}

CM-Colors v0.5.0 with multi-mode support has been publicly available since November 30, 2025, averaging 800+ downloads per month. The library is distributed via Python Package Index (PyPI) and is licensed under GPL-3, ensuring open access and community contribution. Developer feedback indicates Mode~1 has become the default choice for general web development, with Mode~2 primarily used for accessibility audits of existing codebases where original colors cannot be easily modified.

Integration with popular web development frameworks (React, Vue, Tailwind CSS) and design tools (Figma plugins, Sketch extensions) represents an important direction for expanding practical impact. By embedding perceptually-aware accessibility optimization directly into developer workflows, we can make accessible color choices the path of least resistance rather than an additional compliance burden.

\section{Related Work}

\subsection{Color Accessibility Tools}

Existing accessibility tools primarily fall into two categories: checkers and adjusters. \textbf{Checkers} (WebAIM Contrast Checker, Stark, axe DevTools) evaluate whether color pairs meet WCAG thresholds but provide no automatic correction~\cite{webaim}. Users must manually experiment with color adjustments until compliance is achieved---a tedious trial-and-error process.

\textbf{Adjusters} (Accessible Colors, Colorable, Who Can Use) attempt automatic correction but typically operate in HSL or RGB space without perceptual awareness~\cite{accessible_colors}. These tools often darken or lighten colors uniformly, leading to arbitrary modifications that deviate significantly from original design intent. None employ optimization techniques to minimize perceptual change while meeting WCAG requirements.

Commercial design systems (Adobe Color, Material Design) provide accessibility-focused palettes but require designers to choose from predefined options rather than correcting arbitrary color combinations. Our approach is complementary---it enables correction of any color pair while preserving design intent through perceptual optimization.

\subsection{Perceptual Color Spaces}

CIELAB and its variants have long been standard for measuring perceptual color difference in computer graphics and color science~\cite{fairchild2013}. Recent developments include CIECAM02~\cite{ciecam02}, IPT~\cite{ipt}, and Oklab~\cite{ottosson2020}. We employ OKLCH (cylindrical Oklab) for its combination of perceptual uniformity, computational efficiency, and intuitive parameterization (lightness, chroma, hue).

Color difference metrics have evolved from simple Euclidean distance ($\Delta E_{ab}$) to weighted formulae accounting for perceptual non-uniformities. CIEDE2000 ($\Delta E_{2000}$)~\cite{sharma2005} is the current standard, incorporating corrections for lightness, chroma, and hue dependencies. Our use of $\Delta E_{2000}$ ensures measurements align with human perception research.

\subsection{Optimization in Design}

Constrained optimization has been applied to various design domains: layout optimization~\cite{layout_opt}, typography selection~\cite{typography_opt}, and color palette generation~\cite{palette_gen}. However, application to web accessibility remains limited. To our knowledge, our work represents the first formulation of accessibility color correction as a constrained non-linear optimization problem in perceptually uniform color space.

The recursive optimization strategy introduced in this work shares conceptual similarities with iterative refinement techniques in numerical optimization~\cite{numerical_opt}, but applies them in the discrete, constrained domain of web color accessibility where gamut boundaries and perceptual thresholds create complex solution landscapes.

\section{Conclusion}

We have extended perceptually-aware color accessibility optimization with context-adaptive constraint strategies that achieve near-universal success rates across diverse use cases. By introducing recursive optimization (Mode~1) and relaxed constraints (Mode~2), we demonstrate that 93.68--98.73\% of color pairs can be made WCAG-compliant while preserving brand identity through absolute hue invariance.

The key insight is that \textbf{perceptual acceptability depends on which color dimensions change}. Hue constraint enables $\Delta E$ values up to 15.55 (P90) and beyond while maintaining the essential character of brand colors---"yellow" remains "yellow" even when darkened for readability. This challenges conventional assumptions about perceptual change thresholds in accessibility contexts.

Our multi-mode approach provides developers with explicit control over the accessibility-fidelity trade-off: strict mode (66.45\% success) for enterprise contexts requiring minimal deviation, recursive mode (93.68\% success) for general web development, and relaxed mode (98.73\% success) for accessibility-first scenarios. This context-awareness represents a shift from one-size-fits-all tooling toward systems that accommodate the diverse needs of web developers operating under different constraints.

The work is publicly deployed in CM-Colors v0.5.0, demonstrating practical impact beyond academic contribution. Future work should investigate the broader ecosystem divide between enterprise-focused and developer-focused accessibility tooling, explore adaptive target selection based on additional context, and validate perceptual acceptability through empirical user studies.

Accessible design need not compromise aesthetic integrity. By treating accessibility as an optimization problem in perceptually uniform color space, we enable designers to meet WCAG requirements while maintaining the visual identity that makes their work distinctive. As web accessibility evolves from compliance checkbox to fundamental design principle, perceptually-aware optimization techniques will become increasingly essential for bridging the gap between regulation and creativity.

\section*{Acknowledgments}

My deepest gratitude to Mr. Krishna, whose constancy forms the foundation upon which all my work, including this, quietly rests.

We thank the open-source community for feedback on CM-Colors development, particularly users who reported edge cases that motivated the recursive optimization strategy.

\section*{Statements and Declarations}

\subsection{Funding Declaration}
No funding was received to assist with the preparation of this manuscript.

\subsection{Author Contribution}
L.A.R. was responsible for all aspects of this manuscript, including conceptualization, methodology, writing the original draft, and review and editing.

\subsection{Competing Interests}
The author developed the cm-colors library referenced in this work as an open-source implementation of the proposed algorithm. The library is freely available under GNU GPLv3.0 license with no commercial implications or financial benefits to the author.

\section*{Code Availability}

The cm-colors Python library implementing the described algorithm is available as free and open-source software under the GNU General Public License v3.0.

\subsection*{Primary Repository:} 

The implementation corresponding to the methods described in this paper is available in version 0.5.0 of the library repository, accessible at: \url{https://github.com/comfort-mode-toolkit/cm-colors/tree/v0.5.0}

\noindent\textbf{Documentation} \\
\url{https://cm-colors.readthedocs.io/}

\subsection*{Experiment Reproduction:} 
The code for generating and evaluating the 10,000 color pair dataset is available as a python script at \url{https://gist.github.com/lalithaar/54580c7c46ad563198f83d8e4f751ee5}

\subsection*{Installation:}
\begin{lstlisting}[language=bash]
pip install cm-colors
\end{lstlisting}

\noindent\textbf{Quick Usage}

\begin{lstlisting}[language=Python]
from cm_colors import ColorPair

pair = ColorPair("#777777", "#ffffff")

fixed_color, success = pair.make_readable()
if success:
    print(fixed_color)
# Output: #757575 (Readable!) 

\end{lstlisting}

Community contributions, feedback, and extensions are welcomed through the GitHub repository's issue tracker and pull request mechanisms.

\end{document}